\documentclass[12pt]{iopart}

\usepackage{iopams}
\usepackage{graphicx}

\begin{document}

\title{Energy landscape and conical intersection points of the driven Rabi model}

\author{Murray T. Batchelor$^{1,2,3}$, Zi-Min Li$^{1}$ and Huan-Qiang Zhou$^{1}$}

\address{$^{1}$Centre for Modern Physics, Chongqing University, Chongqing 400044, China}

\address{$^{2}$Department of Theoretical Physics, 
Research School of Physics and Engineering, Australian National University, Canberra, ACT 0200, Australia}

\address{$^{3}$Mathematical Sciences Institute, Australian National University, Canberra ACT 0200, Australia}

\ead{batchelor@cqu.edu.cn}

\begin{abstract}
We examine the energy surfaces of the driven Rabi model, 
also known as the biased or generalised Rabi model, as a function of the coupling strength and the driving term.
The energy surfaces are plotted numerically from the known analytic solution.
The resulting energy landscape consists of an infinite stack of sheets connected by conical intersection points 
located at the degenerate Juddian points in the eigenspectrum. 
Trajectories encircling these points are expected to exhibit a nonzero geometric phase. 
\end{abstract}

The driven, biased or generalised Rabi model has hamiltonian \cite{Braak,Chen,Larson}
\begin{equation}
H=\omega \, a^{\dagger}a + g \, \sigma_x(a^{\dagger}+a)+ \Delta \, 
\sigma_z+\epsilon\,\sigma_x \,, \label{ham}
\end{equation}
where $\sigma_x$ and $\sigma_z$ are Pauli matrices for a two-level system with level splitting $\Delta$.
The interaction between the spin and the single-mode bosonic field of frequency $\omega$ is via the coupling $g$. 
The bosonic creation and destruction operators $a^\dagger$ and $a$ satisfy $[a, a^\dagger] = 1$.
The Rabi model is well known as arguably the simplest model for light interacting with matter.
As such, it has a long history in quantum optics.\footnote{
For realisations in ion traps and in both cavity and circuit quantum electrodynamics, 
the reader is referred to Ref. \cite{QED} and references therein.}
The addition of the driving or bias term $\epsilon \, \sigma_x$ breaks a $Z_2$ symmetry (parity).
This additional term allows tunnelling between the two atomic states. 
The driven Rabi model (\ref{ham}) is  
relevant to the description of various hybrid mechanical systems \cite{hybrid,Heun2}. 
In particular, of a micromechanical resonator via coupling to a Cooper-pair box \cite{hybrid,Heun2,ABS,IS}.

Using the analytic solution obtained by Braak \cite{Braak} for the energy eigenspectrum, 
we explore the energy levels of this model as a function of the parameters $g$ and $\epsilon$. 
Specifically, the $N$th eigenvalue is given by $E_N = x_N - g^2/\omega$, where $x_N$  is the $N$th zero of 
\begin{equation}
G_\epsilon(x) = \Delta^2  \bar{R}^+(x) \bar{R}^-(x) - R^+(x) R^-(x) \,, 
\end{equation}
where
\begin{eqnarray}
R^\pm(x)&=& \sum_{n=0}^\infty K_n^\pm(x)  \left( \frac{g}{\omega} \right)^n , \\
\bar{R}^\pm(x)&=& \sum_{n=0}^\infty \frac{K_n^\pm(x)} {x - n \, \omega \pm \epsilon} \left( \frac{g}{\omega} \right)^n .
\end{eqnarray}
$K_n^\pm(x)$ is defined recursively by
$n K_n^\pm = f_{n-1}^\pm(x) \, K_{n-1}^\pm  - K_{n-2}^\pm$ with initial conditions $K_0^\pm=1, K_1^\pm(x)=f_0^\pm(x)$, and 
\begin{equation}
f_n^\pm(x)  = \frac{2g}{\omega} + \frac{1}{2g} \left( n \omega - x  \pm \epsilon + \frac{\Delta^2}{x - n\, \omega \pm \epsilon} \right) .
\end{equation}
The function $G_\pm(x)$ is obtained as a consistency condition for two different series expansions in a 
representation of the bosonic creation and annihilation operators in the Bargmann space of analytical functions \cite{Braak}. 
It can also be derived using the extended coherent states approach \cite{Chen} and 
written in terms of confluent Heun functions.
The eigenstates have also been obtained in terms of confluent Heun functions \cite{Heun2,others2}.

\begin{figure}[t]
\begin{center}
\includegraphics[width=0.32\columnwidth]{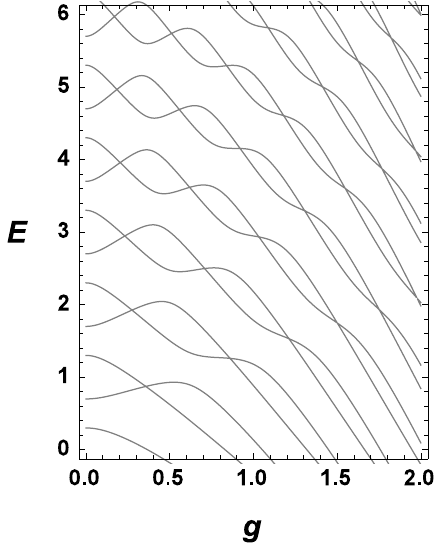}
\includegraphics[width=0.32\columnwidth]{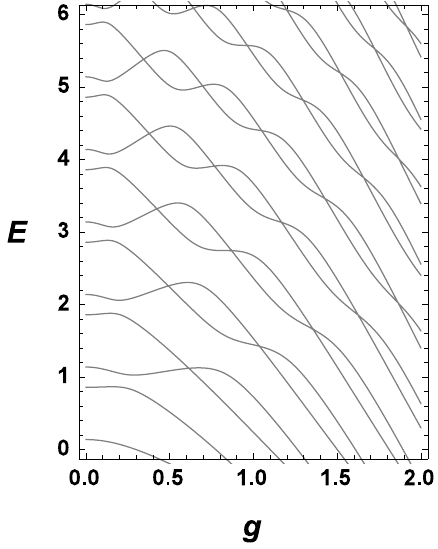}
\includegraphics[width=0.32\columnwidth]{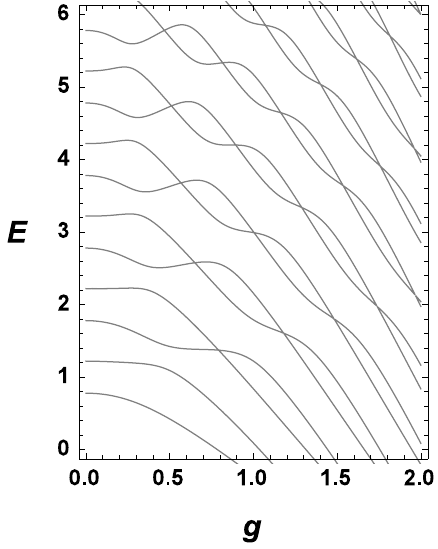}
\caption{Energy eigenspectrum of the driven Rabi model as a function of the coupling $g$ for parameter values $\Delta=0.7$ and $\omega=1$.
From left to right, $\epsilon = 0, \frac12$ and 1.}
\label{fig1}
\end{center}
\end{figure}

The lowest energy levels are shown in Figure 1 as a function of $g$ for fixed $\Delta$ and $\epsilon$.
For $\epsilon=0$ this is the well known plot featuring the Juddian crossing points \cite{Judd} 
for which Kus \cite{Kus} provided a proof that for each value of $N$ there are $N$ crossings if $\Delta$ is in the range $0 < \Delta/\omega < 1$. 
More generally for $k < \Delta/\omega < k+1$ there are $N-k$ crossing points \cite{Kus}.
There are also such crossing points in the eigenspectrum when $\epsilon$ is an integer multiple of $\frac12 \omega$  \cite{Braak,Heun2}.
These are shown in Figure 1 for $\epsilon=\frac12$ and $\epsilon=1$. 
It has also been argued that for a given value of $N$ there are $N$ level crossings for $0 < \Delta/\omega < \sqrt{1+2\epsilon/\omega}$,  
reducing to $N-k$ crossing points for $\Delta$ in the range \cite{LB}\footnote{The eigenspectrum is symmetric in $\epsilon$, 
so this relation holds in general with $\epsilon$ replaced by $|\epsilon|$.}
\begin{equation}
\sqrt{k^2 + 2 k \epsilon/ \omega} < \Delta/\omega < \sqrt{(k+1)^2 + 2(k+1) \epsilon/ \omega} \,.
\label{cross}
\end{equation}

Here we explore the energy landscape as a function of the parameters $g$ and $\epsilon$ for fixed $\Delta$. 
As we shall see, the effect of varying the parameter $\epsilon$ is to induce conical intersection points. 
The tips of the cones are precisely the degenerate Juddian points. 
Figure 2 shows the structure of the interconnected energy surfaces as a function of $g$ and $\epsilon$ for $\Delta=0.7$.
This figure should be viewed in conjunction with Figure 1, which gives the cross-sections at $\epsilon = 0, \pm \frac12, \pm 1$ 
of the energy landscape shown in Figure 2. 
This figure also shows crossing points for $g=0$ as the parameter $\epsilon$ is varied. 
These are determined by the energies
\begin{equation}
E = N \omega \pm \sqrt{\Delta^2 + \epsilon^2} \,.
\end{equation}
Two more refined views of the energy surfaces are shown in Figure 3.\footnote{
Note that the overall structure of these figures can be made simpler by plotting $E + g^2/\omega$ 
on the $z$-axis which has the effect of turning the paraboloids into planes.} 
It is clear that the energy surfaces for $g>0$ are connected by isolated conical intersection points.
These are shown further in Figure 4, which illustrates the elementary cones.

\begin{figure}[t]
\begin{center}
\includegraphics[width=0.8\columnwidth]{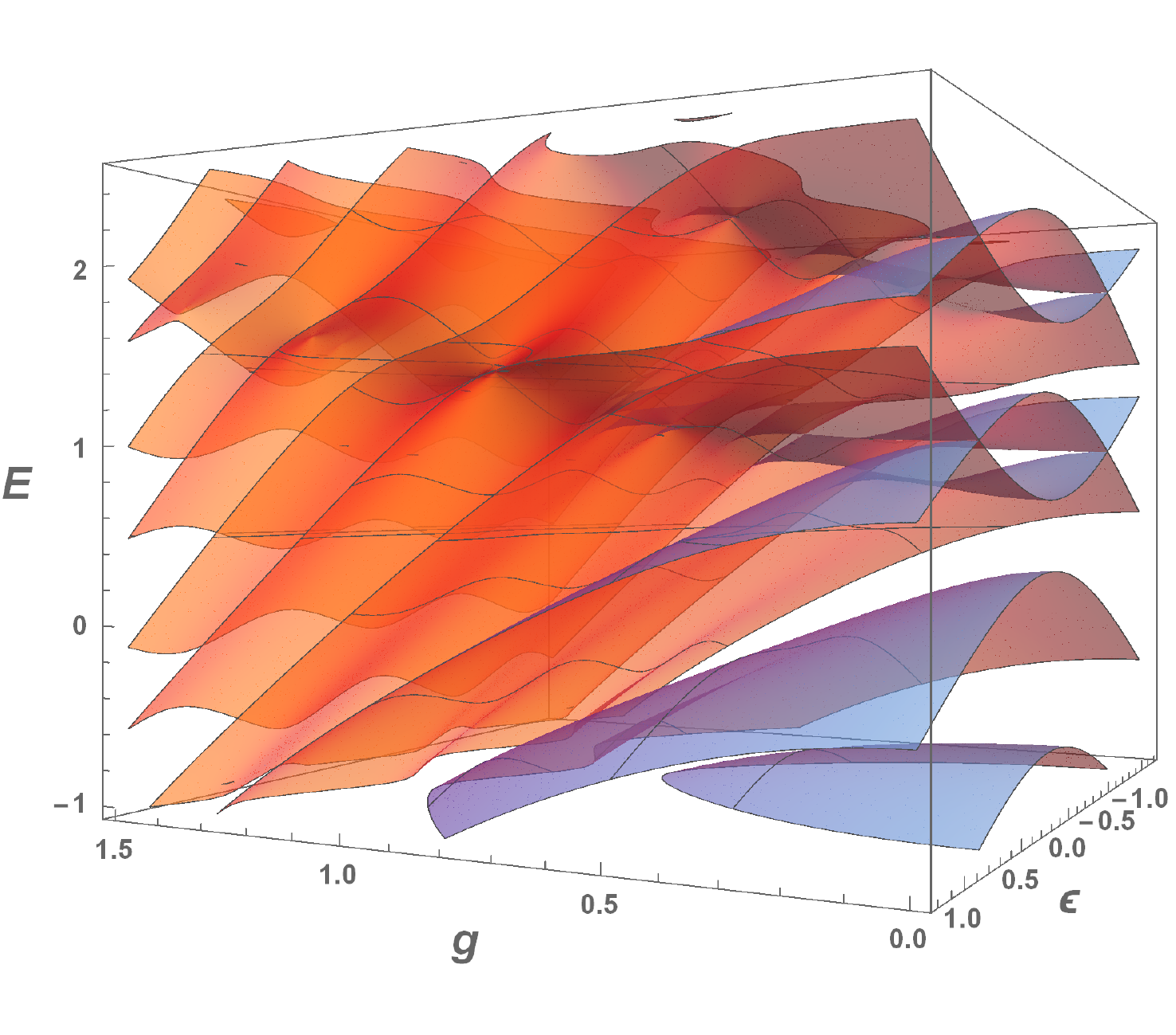}
\caption{Energy landscape of the driven Rabi model as a function of the parameters $g$ 
and $\epsilon$ for the values $\Delta=0.7$ and $\omega=1$. The landscape consists of a series of discrete sheets labelled 
by integer $N$ and connected by conical intersection points for $N \ge 1$. }
\label{fig2}
\end{center}
\end{figure}

\begin{figure}[t]
\begin{center}
\includegraphics[width=0.45\columnwidth]{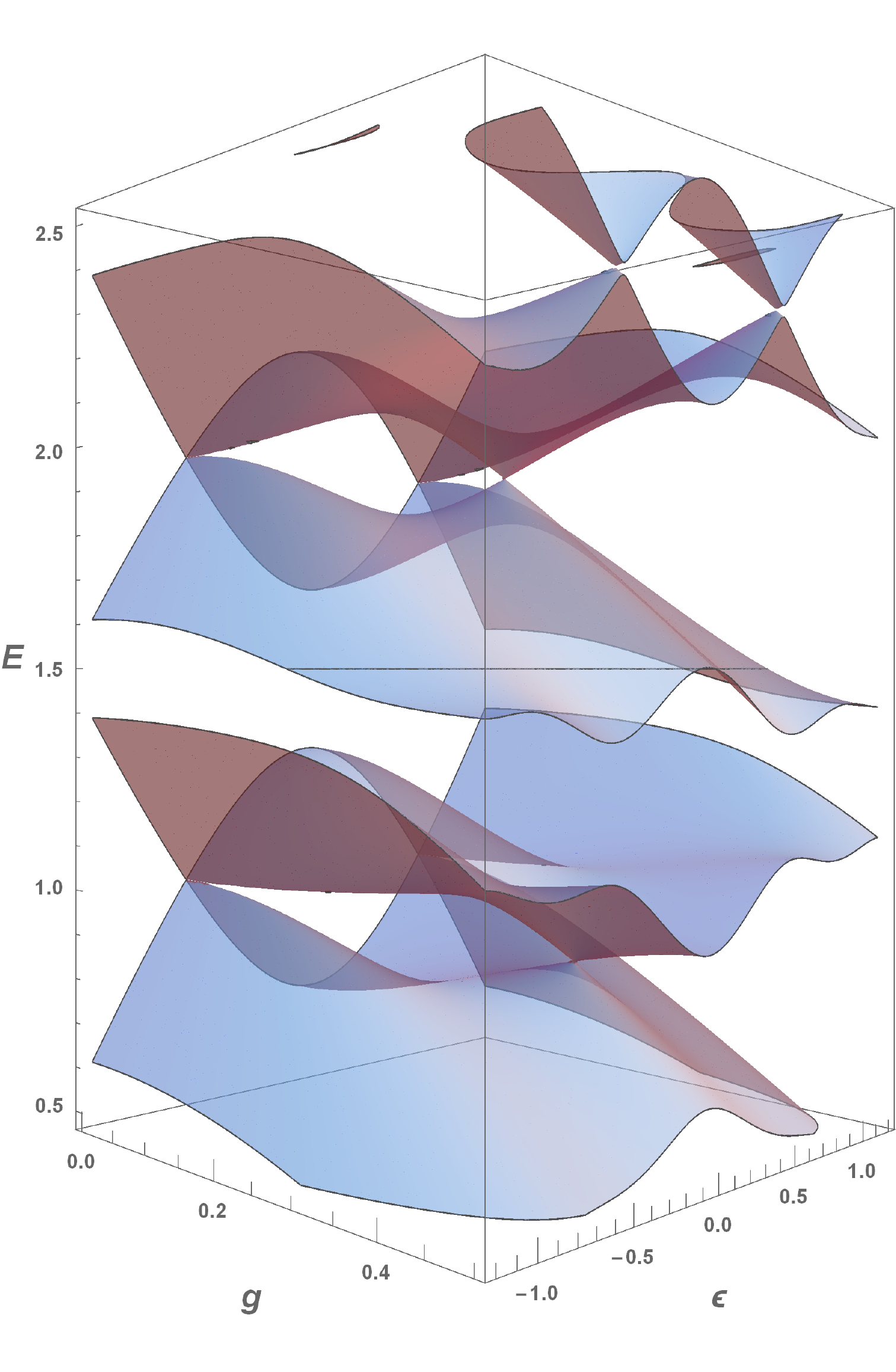}
\includegraphics[width=0.45\columnwidth]{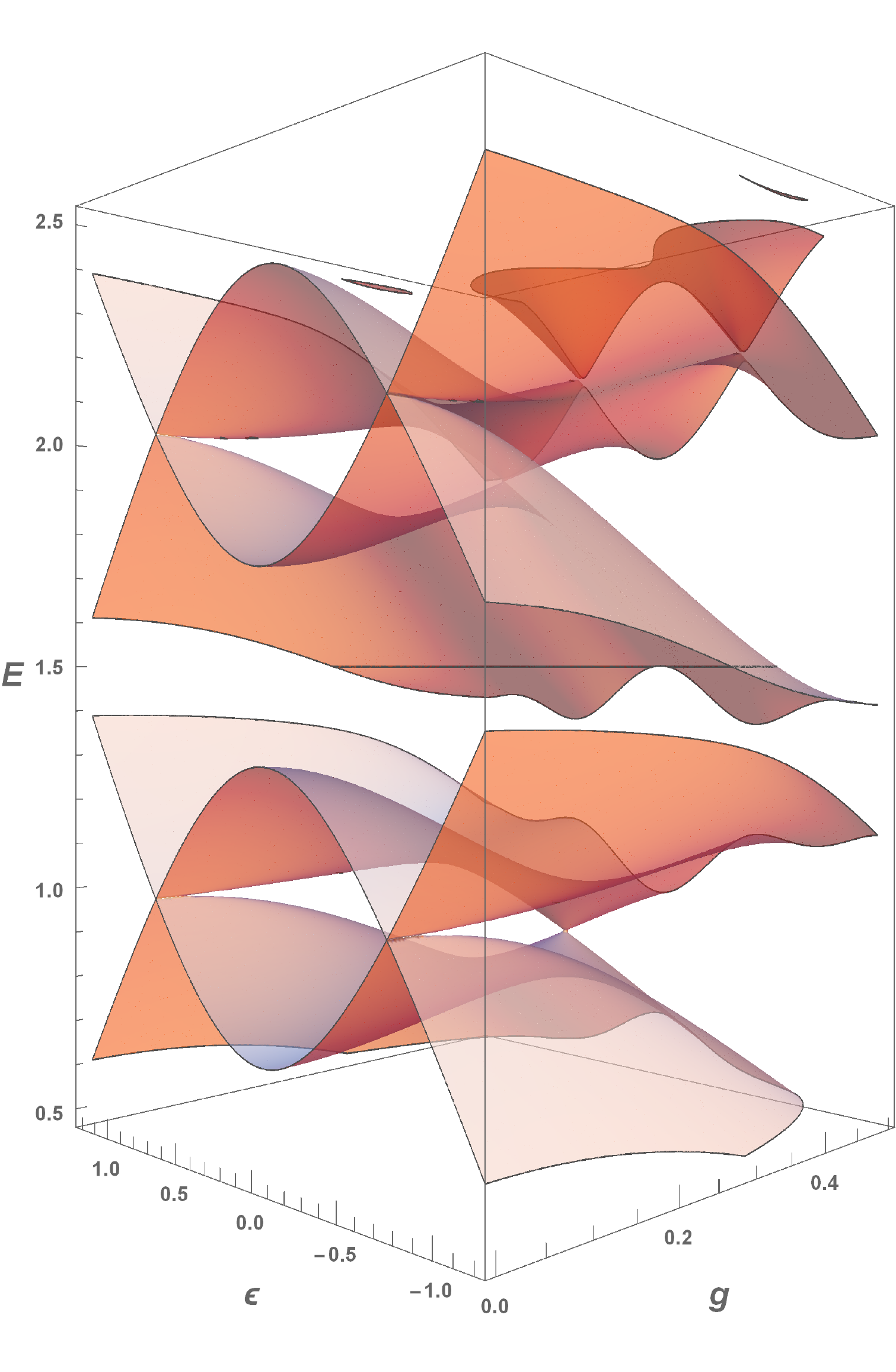}
\caption{Close up views of the energy landscape depicted in Figure 2 showing the energy surfaces connected by 
conical intersection points.}
\label{fig3}
\end{center}
\end{figure}

\begin{figure}[t]
\begin{center}
\includegraphics[width=0.45\columnwidth]{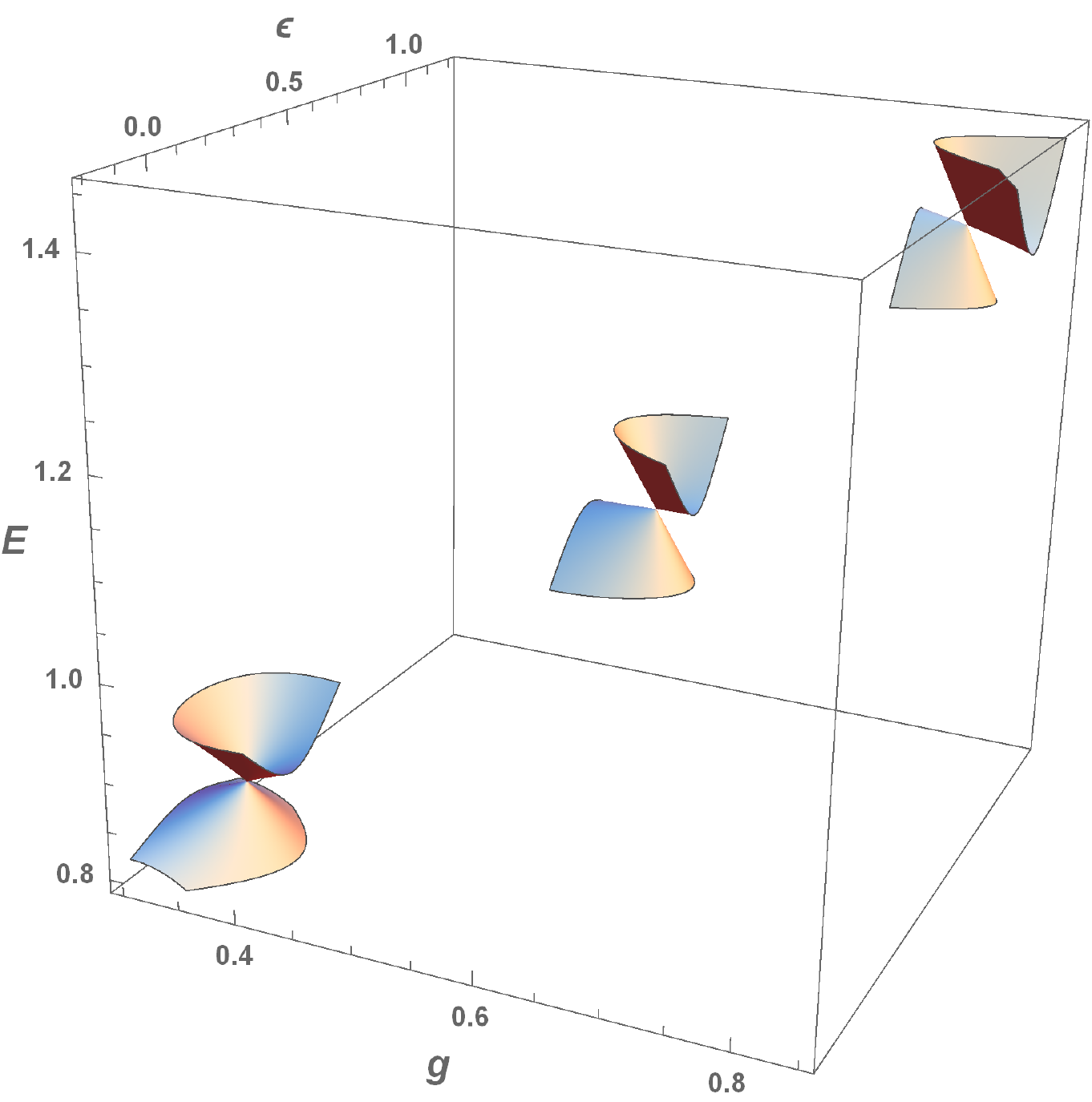}
\includegraphics[width=0.45\columnwidth]{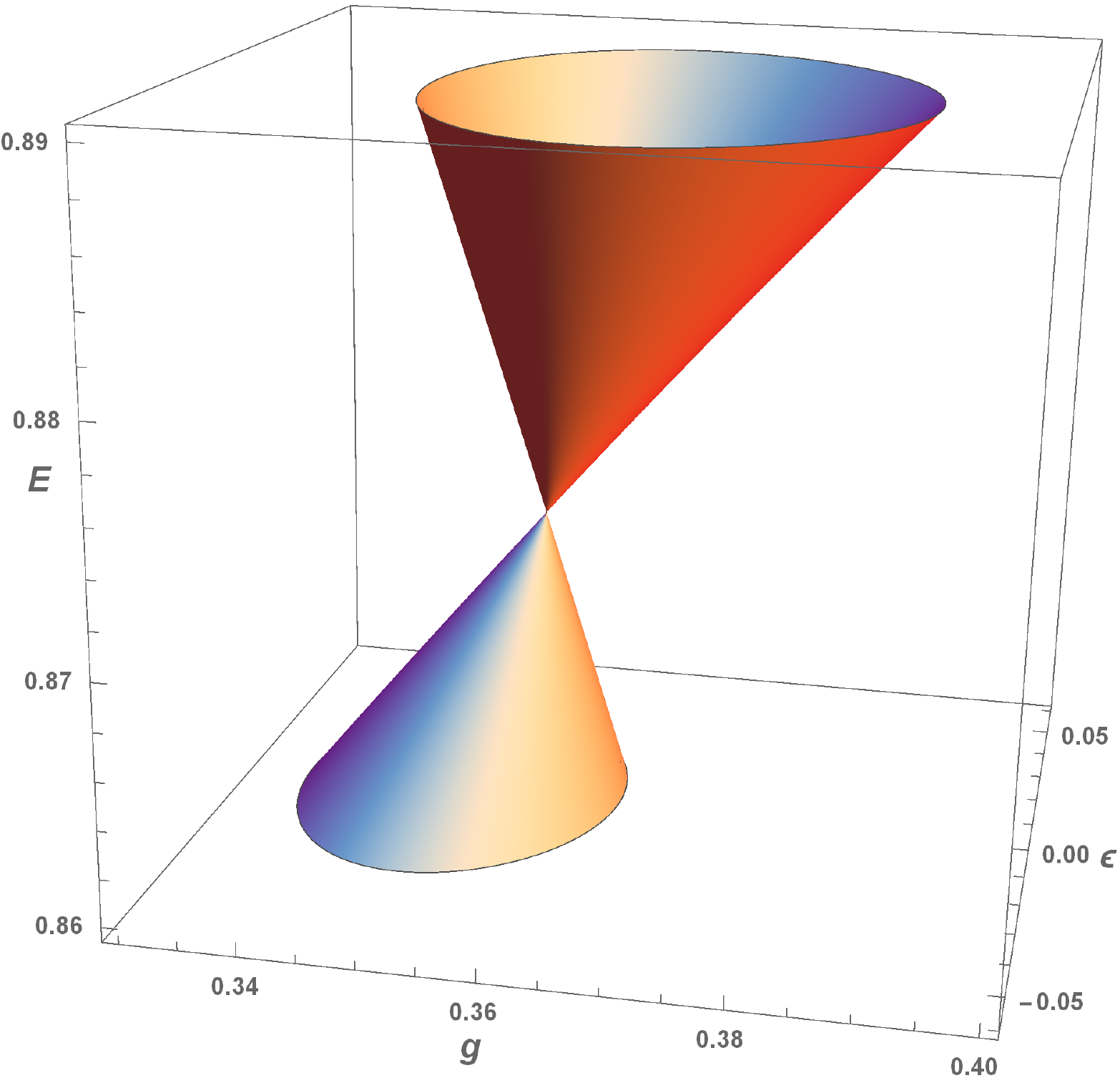}
\caption{Elementary cones in the energy spectrum for $\Delta=0.7$ with $\omega=1$. 
The plot on the left shows the ``lowest'' 
cones at $\epsilon=0, \frac12$ and $1$. Also shown is a magnification of the cone 
centred at $\epsilon=0$.}
\label{fig4}
\end{center}
\end{figure}

The overall structure of the energy landscape is an infinite stack of sheets labelled by integers $N=0,1,2,\ldots$ 
connected for $g > 0$ by conical intersection points for $N \ge 1$.
The conical intersection points are the exceptional points in the eigenspectrum which are located on lines 
in the planes defined by $\epsilon= \pm \frac12 n \omega$, $n=0,1,2,\ldots$.
These lines are the intersections of the surfaces $E=N_1 \omega - g^2/\omega - \epsilon$ and 
$E=N_2 \omega - g^2/\omega + \epsilon$ for different $N_1, N_2 \ge 1$ with $N_1-N_2 = 2 \epsilon/\omega$.
The precise locations of the conical intersection points along these lines are determined by the solutions of the 
constraint relations \cite{Heun2,LB}.
Their number is dependent on the value of $\Delta/\omega$.
From the above consideration of the number of crossing points, 
the number of conical intersection points can be tabulated for given $\Delta/\omega$ and $\epsilon/\omega$.
In this sense the precise geometry of the energy landscape is thus known.
For example, for the value $\Delta=0.7$ considered in the figures, there are $N$ conical intersection points 
for a given value of $N$ located in each of the planes where $\epsilon= \pm \frac12 n \omega$, $n=0,1,2,\ldots$. 
Figure 4 identifies the most elementary of the cones, with a magnification of the cone centred at $\epsilon=0$.
From the general result (\ref{cross}) for the number of crossing points we know that 
this particular cone vanishes for $\Delta/\omega > 1$.

Trajectories encircling conical intersection points are expected to exhibit a nonzero geometric phase. 
Geometric phases \cite{Berry,Book} have been investigated in the context of the Rabi model for some time, 
and not without debate (see, e.g., Refs. \cite{Larson2,Deng,Wang} and references therein).
For systems like the Rabi model the Berry phase is induced by integration over a unitary transformation parameter $\varphi$ 
with the angle variable $\varphi$ varying slowly from $0$ to $2\pi$ \cite{phase}.
Here we have demonstrated conical intersection points in the energy landscape 
for the {driven} Rabi model by varying the system parameters $g$ and $\epsilon$.
Since analytic expressions for the corresponding eigenstates are known for this model, 
the geometric phases obtained by integrating trajectories encircling the conical points 
should, in principle, be calculable.

As a final remark, we note that the original justification for studying Juddian points was that, 
because they were exact solutions, they could be useful in testing and improving various approximation schemes. 
Here, using the analytic solution of the driven Rabi model, 
we have demonstrated that the degenerate Juddian points are the conical intersection points 
in the energy landscape.
Degenerate Juddian points exist in a range of models related to the Rabi model.
It is hoped that these models can be added to the different contexts in which conical intersection points 
have been observed.\footnote{See, e.g., Refs. \cite{laser,gas,nature} and references therein.}
The conical intersection points are expected to play a crucial role in the dynamics of the driven Rabi model, 
which is as yet only partially \cite{Larson} explored.

\ack
It is a pleasure to thank Michael Berry for helping to initiate this work during his visit to 
Chongqing University and the Qiantang River in September 2014.
MTB gratefully acknowledges support from Chongqing University and the 1000 Talents Program of China. 
This work has also been partially supported by the National Natural Science Foundation of China (Grant No.~11174375) 
and the Australian Research Council through grant DP130102839.

\end{document}